\documentclass[prb,twocolumn,floatfix,showpacs,superscriptaddress]{revtex4-1}
\usepackage{graphicx,epsfig,color}
\usepackage[ 
    colorlinks=true,
    pdfborder={0 0 0},
    allcolors=blue,
    breaklinks=true
]{hyperref}

\usepackage{threeparttable}
\usepackage{breakcites}
\usepackage[utf8]{inputenc}
\begin{document}

\title{Comparative Investigation of MAPbBr$_x$I$_{1-x}$ and SbH$_4$Br$_x$I$_{1-x}$ Perovskites: Electronic and Structural Properties}

\author{Veysel \c{C}elik}\email{veysel3@gmail.com}
\affiliation{Department of Mathematics and Science Education, Siirt University, Siirt 56100, Turkey}

\date{\today}

\begin{abstract}
This paper comparatively investigates the structural and electronic properties of hybrid perovskites MAPbBr$_x$I$_{1-x}$ and SbH$_4$PbBr$_x$I$_{1-x}$ by means of DFT-based calculations. The main aim is to check if the increase in band gap due to substitution of I$^-$ ions with Br$^-$ ions can be overcome by introducing the inorganic SbH$_4^+$ cation. Since the Br$^-$ ions merely enhance structural stability of the perovskite framework and SbH$_4^+$ not only sustains that stability but also reduces the band gap to nearly ideal values and thereby improves electronic performance, these are the leading candidates among them. Of them, the reduced band gap SbH$_4$PbI$_3$ ($\sim$1.37 eV) and the perfectly matched SbH$_4$PbBrI$_2$ with its band gap being exactly 1.51 eV are top prospects for being stable and having high-efficiency solar cell applications. The findings show that SbH$_4^+$-based perovskites have potential for future photovoltaic devices.

\end{abstract}


\maketitle
                                    
\section{Introduction}

Among photovoltaic researches, methylammonium lead iodide (MAPbI$_3$) is the most researched perovskite compound. This compound has attracted great attention as it possesses near-ideal band gap of approximately 1.55 eV that facilitates the effective absorption of a broad solar spectrum and enables high photocurrent generation~\cite{eperon2014, noh2013}. Its desirable optoelectronic properties and crystal structure have been key determinants of the tremendous progress realized in the power conversion efficiencies (PCEs) of perovskite solar cells in the past ten years~\cite{green2014, NREL2025}.

Hybrid organic–inorganic metal–halide perovskites, and more particularly MAPbI$_3$, have been the subject of intense research due to their exquisite photovoltaic characteristics. Investigation of antimony hydride–based perovskites such as SbH$_4$PbI$_3$ has been relatively sparse. Filip \textit{et al.}~\cite{Filip2014} systematically investigated perovskite compositions by replacing the A-site cations with hydride-based ions (NH$_4^+$, PH$_4^+$, AsH$_4^+$, SbH$_4^+$) and studied the resulting electronic band structures using GW calculations. They discovered that SbH$_4$PbI$_3$ possesses a convenient direct band gap of approximately 1.4 eV, which is suitable for photovoltaic applications. Furthermore, these materials possess small electron and hole effective masses (less than 0.3 electron masses), indicating good charge transport properties~\cite{Filip2014, Filip2015}. The electronic properties of SbH$_4$PbI$_3$ are therefore highly promising.

But to look into the electronic properties of SbH$_4$PbI$_3$, along with its stability it is also important. Tenuta \textit{et al.}~\cite{Tenuta2016} performed density functional theory (DFT) calculations in order to look into the thermodynamic stability of hybrid halide perovskites, SbH$_4$PbI$_3$ being one of them. Their overall conclusion was that the materials have poor thermodynamic stability largely due to inherent structural attributes and environmental susceptibilities. In particular, they identified that chemical stability of SbH$_4$PbI$_3$ is directly related to aqueous solubility of its decomposition product, SbH$_4$I, in a way that the material becomes susceptible to degradation upon exposure to humid environments. Another study, nonetheless, by Rego \textit{et al.}~\cite{Rego2025} more recently showed that the incorporation of van der Waals (vdW) corrections alters drastically the stability predictions, in a way that SbH$_4$PbI$_3$ is thermodynamically favored. Their work also showed that the incorporation of non-local correlation effects via vdW interactions helps remove the discrepancies in standard DFT calculations and hence provides a better estimate of SbH$_4$PbI$_3$ stability. These are theoretical investigations and need more experimental data to back them up. One very certain prediction can be made, however, that the incorporation of the PbI$_3^-$ complex ion will lower the overall stability.

MAPbI$_3$ is known to have severe disadvantages that disqualify it from being practically utilized. In particular, the material is inherently unstable in the presence of moisture, oxygen, and heat stress, which leads to rapid degradation of the photovoltaic performance~\cite{niu2015, wang2019}. Substitution of iodide (I$^-$) in the perovskite structure with bromide (Br$^-$) has been recognized as a probable choice to inhibit these disadvantages. Bromide ions, being smaller in ionic radius than iodide ions, result in the cage parameters decreasing and contribute to greater structural stability~\cite{noh2013, saliba2016}. The enhanced thermal and moisture stability of the resultant perovskite material with stronger Pb–Br bonds, however, has the drawback of a broader band gap reducing absorption of the solar spectrum. Consequently, mixed-halide perovskites comprising both bromide and iodide ions are of intense focus as a method of maximizing efficiency and stability~\cite{abate2014, bush2017}. Moreover, Zheng and Rubel~\cite{Zheng2017} demonstrated that the strong electron affinity of the PbI$_3^-$ complex ion significantly detracts from the thermodynamic stability of perovskites. Therefore, stepwise substitution of iodide atoms by bromide atoms in SbH$_4$PbI$_3$ is suggested as a probable means of reducing electron affinity and thus enhancing structural stability.

In the present work, a strategy consisting of controlled replacement of I$^-$ ions by Br$^-$ ions in fixed ratios was followed. This method can serve as an efficient tool for improving the stability of the material. No results from the literature were found for studies of the SbH$_4$Br$_x$I$_{1-x}$ perovskite structure in which bromide ions are replaced stepwise by iodide ions. Therefore, results of this research can be of primary importance.

The main purpose of this study is to investigate whether the band gap value, which increases as a result of the replacement of I$^-$ ions with Br$^-$ ions, can be reduced by the SbH$_4^+$ cation. In the current research, first-principles density functional theory (DFT) calculations, taking into account van der Waals (vdW) interactions used~\cite{hohenberg1964inhomogeneous, kohn1965self, giannozzi2009quantum, giannozzi2017advanced}. The vdW corrections need to be included to properly mimic the interactions at organo–inorganic interfaces and examine the structural behavior of hybrid perovskites. By including vdW effects, the DFT calculations performed in this study agree more closely with experimental measurements, which increases the reliability of the theoretical method for prediction of photovoltaic material performance.

\section{Computational Methods}

In this study, first-principles calculations were performed to investigate the electronic and structural properties of the material. The calculations were conducted within the framework of density functional theory (DFT)~\cite{hohenberg1964inhomogeneous, kohn1965self} using the Quantum ESPRESSO package~\cite{giannozzi2009quantum, giannozzi2017advanced}. The exchange-correlation effects were treated using the generalized gradient approximation (GGA) with the Perdew-Burke-Ernzerhof (PBE) functional~\cite{perdew1996generalized}. The projector augmented wave (PAW) method~\cite{blochl1994projector, kresse1999ultrasoft} was employed to describe the core-valence electron interactions, with pseudopotentials taken from the PBE PAW pseudopotential library. A variable cell relaxation (\texttt{vc-relax}) calculation was performed to optimize atomic positions and cell parameters. The convergence threshold for the self-consistent field (SCF) calculations was set to $1 \times 10^{-5}$ Ry for the total energy and $1 \times 10^{-4}$ Ry/au for the force convergence. The plane-wave kinetic energy cutoff was chosen as 60 Ry. Brillouin zone integration was carried out using a Monkhorst-Pack~\cite{monkhorst1976special} k-point mesh. A $6 \times 6 \times 6$ k-point grid was used for structural optimization, while a denser $13 \times 13 \times 13$ k-point grid was employed for the density of states (DOS) calculations. To accurately account for van der Waals (vdW) interactions, the DFT-D3 method~\cite{grimme2010consistent} was applied.

The effective mass \( m^* \), which is important for describing the dynamic properties of charge carriers, is obtained from the curvature of the energy band structure. The band structures were obtained by DFT, and the energy dispersion \( E(k) \) near the conduction band minimum or the valence band maximum is approximated as parabolic. This hypothesis enables us to approximate the energy by a second-order Taylor expansion when the wave vector \( k \) is small.

Then, the curvature of the band is analyzed by calculating the second derivative \( \frac{d^2 E}{d k^2} \) within the parabolic range. The effective mass is calculated using the relation

\begin{equation}
m^* = \hbar^2 \left( \frac{d^2 E}{d k^2} \right)^{-1},
\label{eq:effective_mass}
\end{equation}

where \( \hbar \) is the reduced Planck constant. Lastly, effective mass is obtained by multiplying the inverse of curvature by \( \hbar^2 \), where \( \hbar = 1.0545718 \times 10^{-34}~\mathrm{J \cdot s} \) in SI units.

This technique effectively relates the band curvature to the effective mass of charge carriers, therefore providing valuable insights into carrier transport in semiconductor substances.

\section{Results \& Discussion}

\begin{figure}[t]
\includegraphics[width=8cm]{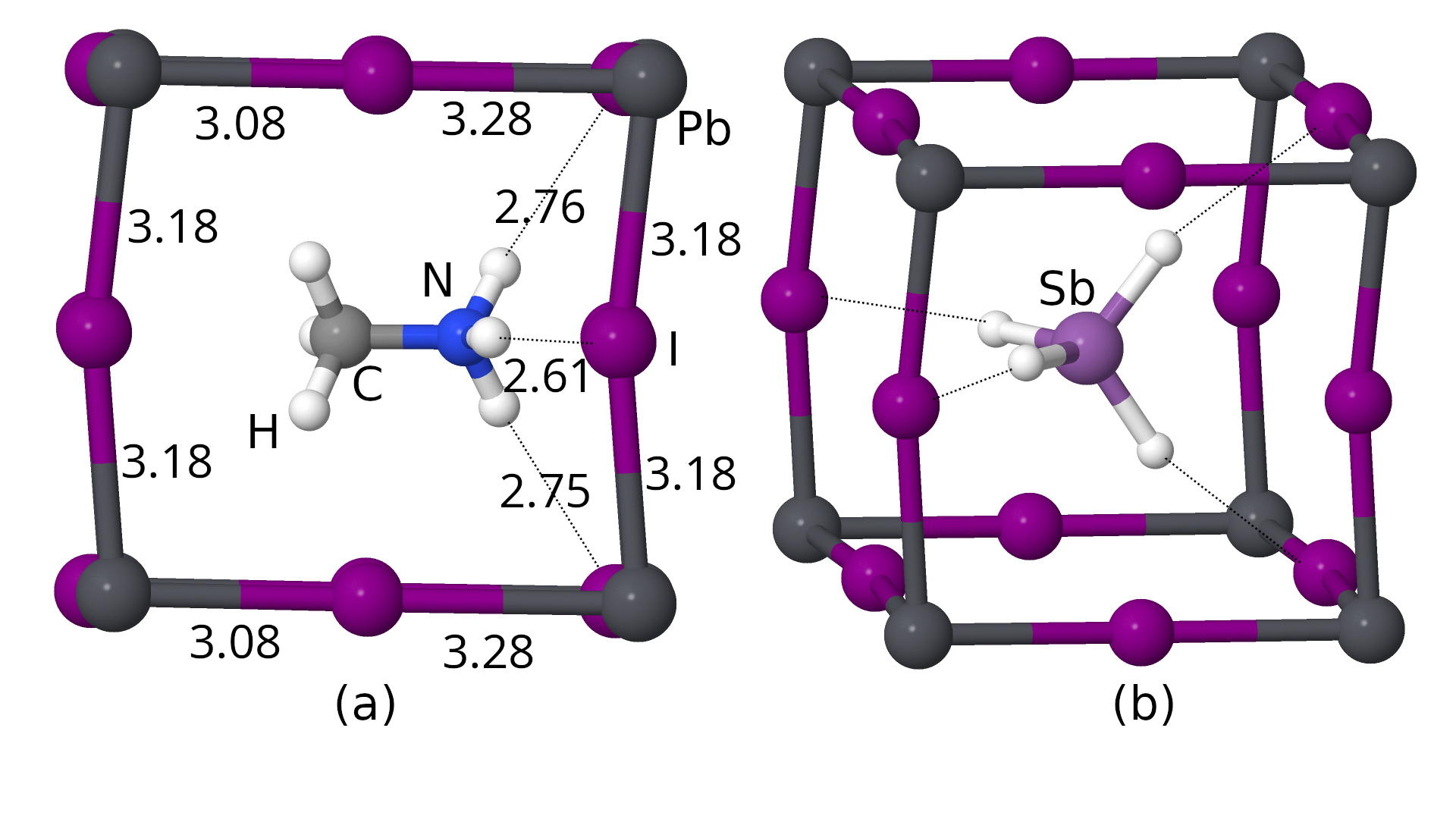}
\caption{The optimized structures of MAPbI$_3$ and SbH$_4$PbI$_3$. In the inorganic lattice, gray and purple balls represent Pb and I ions, respectively. The numbers indicate the lengths of the corresponding bonds in angstroms.}
\label{fig1}
\end{figure}

In this theoretical study, comparison was done by structures derived based on methylammonium lead iodide (MAPbI$_3$), a hybrid organic–inorganic perovskite of special interest after its potential application in solar cells and optoelectronic devices~\cite{stranks2013electron}. The structure parameters derived in the investigated systems are given in Table~\ref{Table1}. For MAPbI$_3$, the computed lattice parameters are $a = 6.36$~\AA, $b = 6.33$~\AA, and $c = 6.37$~\AA. Experimental studies employing X-ray diffraction analysis have reported lattice parameters in the range of 6.32--6.33~\AA~\cite{poglitsch1987dynamic}, in agreement with the calculations presented in this work. As shown in the optimized structure of MAPbI$_3$ presented in Figure~\ref{fig1}, the Pb--I bond lengths on the side where the methylammonium (MA$^+$) cation is attached are bent inward and measured to be 3.18~\AA. This value is in good agreement with the previously reported experimental value of approximately 3.18~\AA~\cite{Schuck2022}. These comparisons indicate that the computational methodology used in this research correctly describes the structural features of MAPbI$_3$.

\begin{table}[htbp]
\caption{Lattice parameters (\textit{a}, \textit{b}, \textit{c}) and arithmetic averages of selected bond lengths (Pb--I, Pb--Br, H--I, H--Br) for the investigated structures. Values are given for both pure and mixed halide compositions.}
\label{Table1}
\centering
\begin{tabular}{|l|c|c|c|c|c|c|c|}
\hline
Structure & \textit{a} & \textit{b} & \textit{c} & Pb--I & Pb--Br & H--I & H--Br \\ \hline
MAPbI$_3$            & 6.36 & 6.33 & 6.37 & 3.22 & --   & 2.71 & --   \\ \hline
SbH$_4$PbI$_3$       & 6.23 & 6.30 & 6.36 & 3.14 & --   & 2.94 & --   \\ \hline
MAPbBr$_3$           & 5.96 & 5.96 & 5.99 & --   & 3.03 & --   & 2.47 \\ \hline
SbH$_4$PbBr$_3$      & 5.84 & 5.91 & 5.96 & --   & 2.96 & --   & 2.75 \\ \hline
MAPbBr$_2$I          & 5.95 & 5.99 & 6.36 & 3.20 & 3.03 & 2.60 & 2.41 \\ \hline
SbH$_4$PbBr$_2$I     & 5.88 & 5.94 & 6.29 & 3.15 & 2.97 & 2.86 & 2.82 \\ \hline
MAPbBrI$_2$          & 5.95 & 6.36 & 6.41 & 3.22 & 3.06 & 2.64 & --   \\ \hline
SbH$_4$PbBrI$_2$     & 5.88 & 6.31 & 6.31 & 3.17 & 2.93 & 2.88 & --   \\ \hline
\end{tabular}
\end{table}

\begin{figure}[t]
\includegraphics[width=8cm]{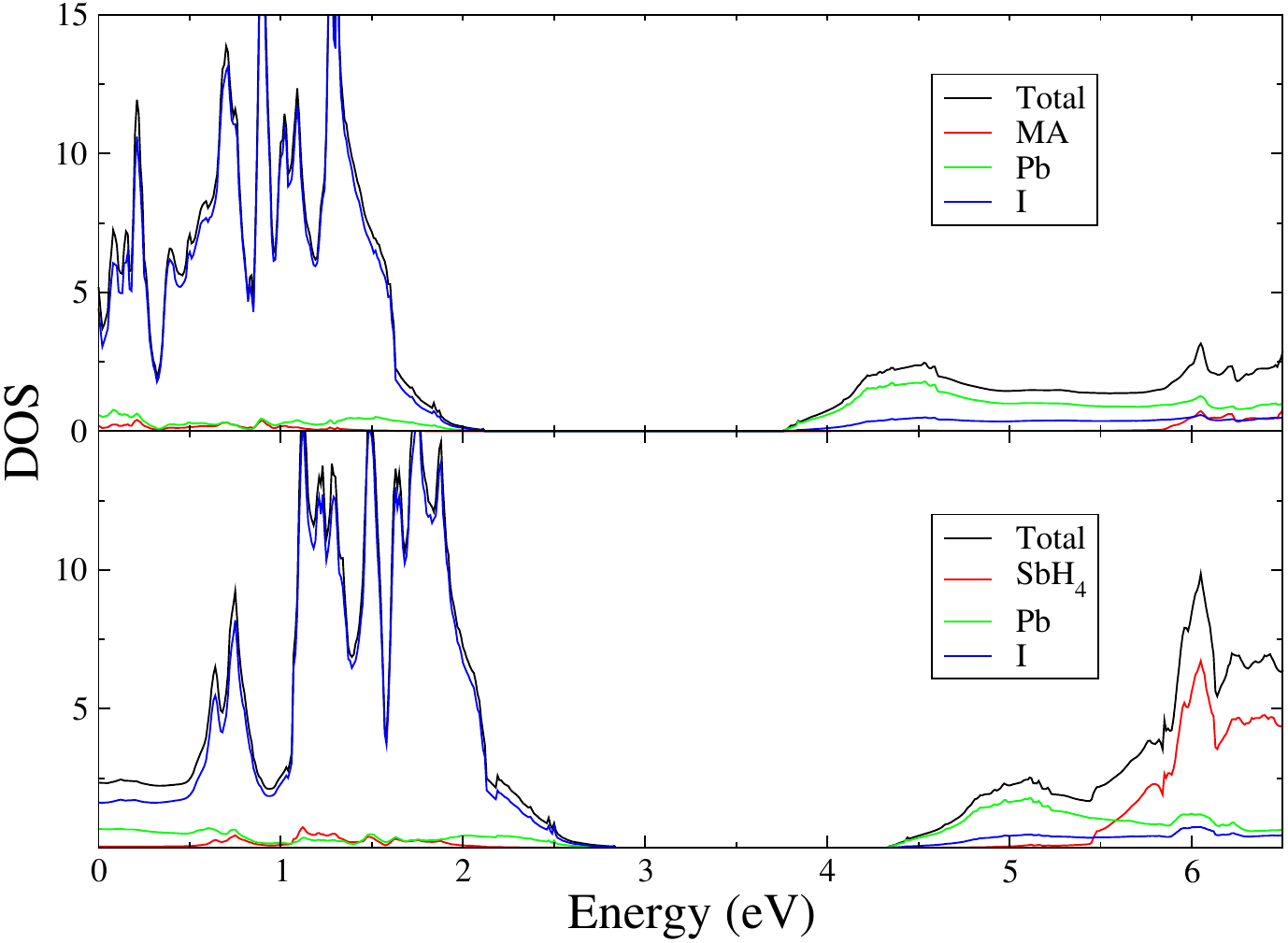}
\caption{Density of States (DOS) for MAPbI$_3$ and SbH$_4$PbI$_3$. The plots display the total DOS along with the individual contributions from the organic cation (MA for MAPbI$_3$ and SbH$_4$ for SbH$_4$PbI$_3$), Pb, and I. Fermi level is not normalized to zero in these plots. Energy is given in electron volts (eV).}
\label{fig2}
\end{figure}

\begin{figure}[t]
\includegraphics[width=8cm]{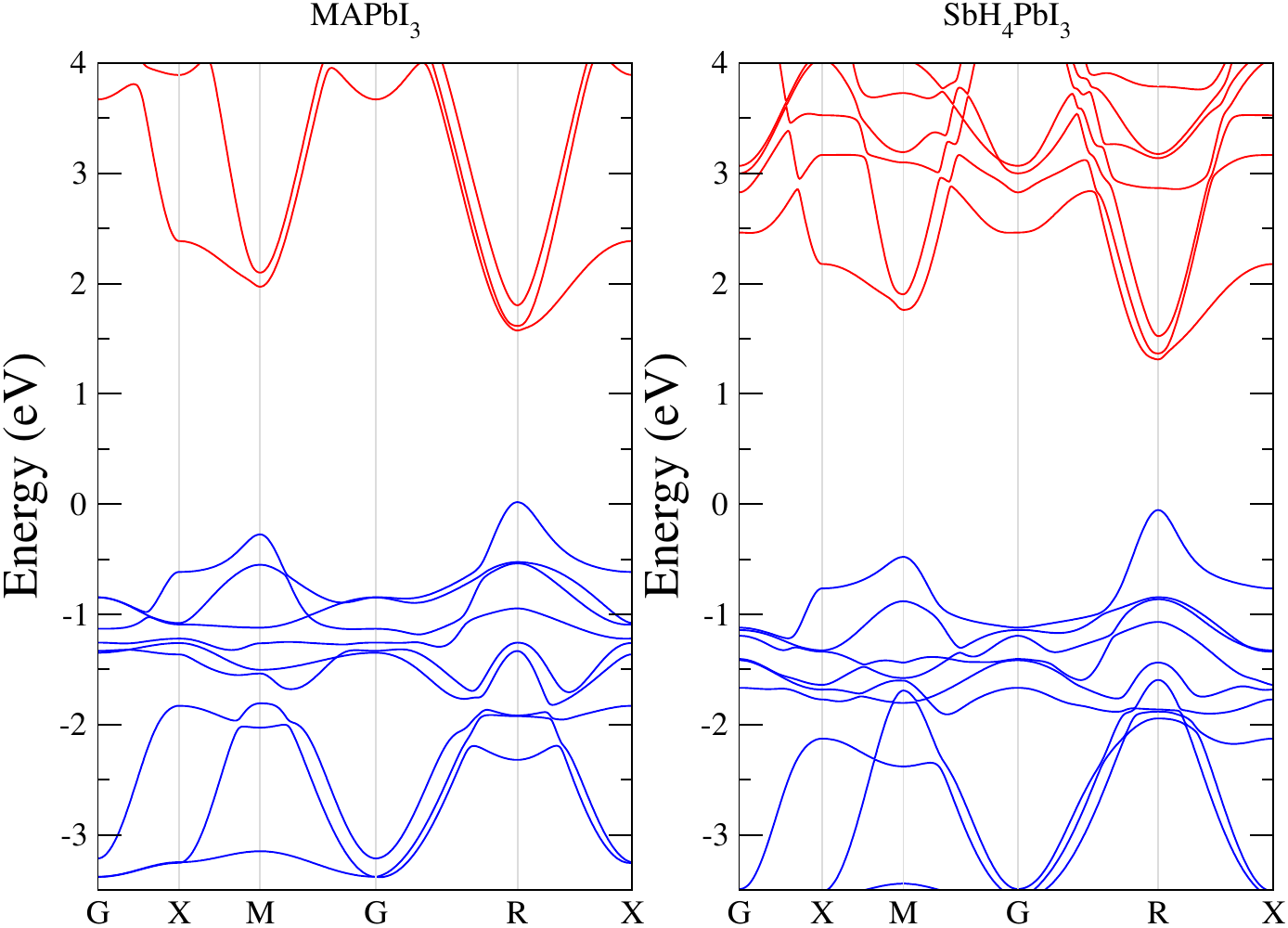}
\caption{Band structures of MAPbI$_3$ and SbH$_4$PbI$_3$ along high-symmetry paths in the Brillouin zone. For these band structure plots, the Fermi level is shifted to zero.}
\label{fig3}
\end{figure}

\begin{table}[t]
\centering
\caption{The table presents the properties of various perovskite materials, including their valence band maximum, conduction band minimum, band gap in electron volts (eV), and effective masses of electrons and holes.}
\label{table2}
\begin{tabular}{lccccc}
\hline
Material & VBM & CBM & $E_g$ (eV) & $m^*_e$ & $m^*_h$ \\
\hline
MAPbI$_3$ & 2.1799 & 3.7387 & 1.5588 & 0.254 & 0.232 \\
MAPbBrI$_2$ & 2.3287 & 3.9798 & 1.6511 & 0.239 & 0.198 \\
MAPbBr$_2$I & 2.5681 & 4.2807 & 1.7126 & 0.170 & 0.194 \\
MAPbBr$_3$ & 2.8141 & 4.7151 & 1.9010 & 0.268 & 0.222 \\
SbH$_4$PbI$_3$ & 2.9328 & 4.3003 & 1.3675 & 0.215 & 0.164 \\
SbH$_4$PbBrI$_2$ & 3.1270 & 4.6416 & 1.5146 & 0.147 & 0.158 \\
SbH$_4$PbBr$_2$I & 3.4400 & 5.0579 & 1.6179 & 0.360 & 0.175 \\
SbH$_4$PbBr$_3$ & 3.6940 & 5.4169 & 1.7229 & 0.289 & 0.182 \\
\hline
\end{tabular}
\end{table}

Figure~\ref{fig1} displays the optimized structures of MAPbI$_3$ and SbH$_4$PbI$_3$. For SbH$_4$PbI$_3$, the computed lattice parameters are $a = 6.23$~\AA, $b = 6.30$~\AA, and $c = 6.36$~\AA. It is observed that, except for a slightly decreased $a$ parameter, the overall lattice size is comparable with that of MAPbI$_3$. The major structural difference between the two perovskite systems stems from the hydrogen-bonding networks. In MAPbI$_3$, the organic MA$^+$ cation forms three N--H$\cdots$I hydrogen bonds with neighboring iodide ions, on average with bond length of approximately 2.71~\AA. The inorganic SbH$_4^+$ cation in SbH$_4$PbI$_3$, however, forms four hydrogen bonds with an average bond length of approximately 2.94~\AA. Additionally, while the MA$^+$ cation will push the iodide ions in towards the PbI$_3^-$ lattice, the SbH$_4^+$ cation likes to bond with the iodide ions on both sides of the lattice, resulting in an outward displacement of the ions. Such a discrepancy in hydrogen-bonding interactions can result in minor changes in lattice parameters and local structural distortions that impact the material's stability and phase behavior~\cite{green2014emergence, kojima2009organometal}. The strengthened hydrogen-bond network of SbH$_4$PbI$_3$ can help with the increased structural stability without producing extensive lattice distortions.

\begin{figure}[h]
\includegraphics[width=8cm]{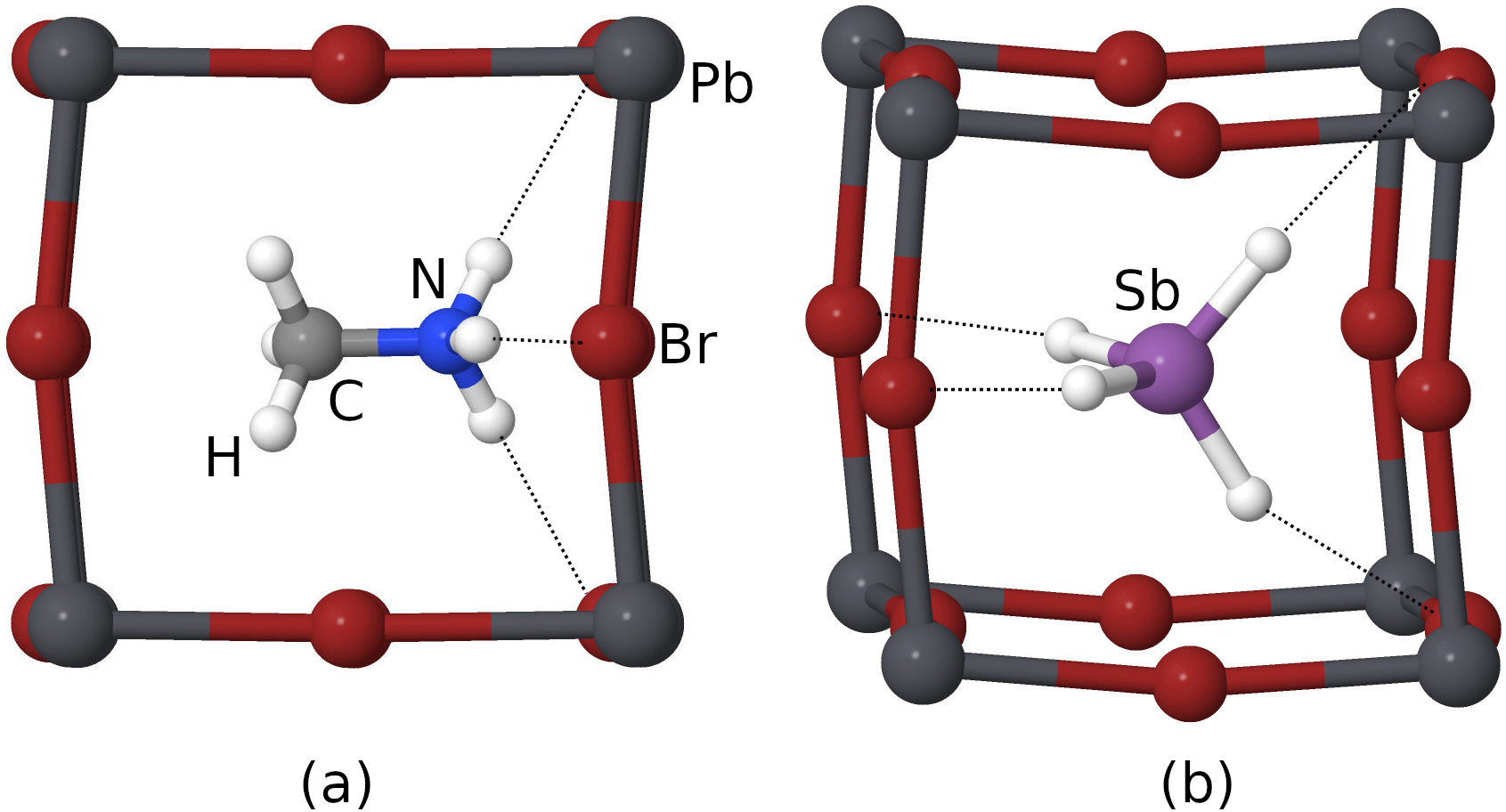}
\caption{The optimized structures of MAPbBr$_3$ and SbH$_4$PbBr$_3$. In the inorganic lattice, gray and red balls represent Pb and Br ions, respectively.}
\label{fig4}
\end{figure}

\begin{figure}[h]
\includegraphics[width=8cm]{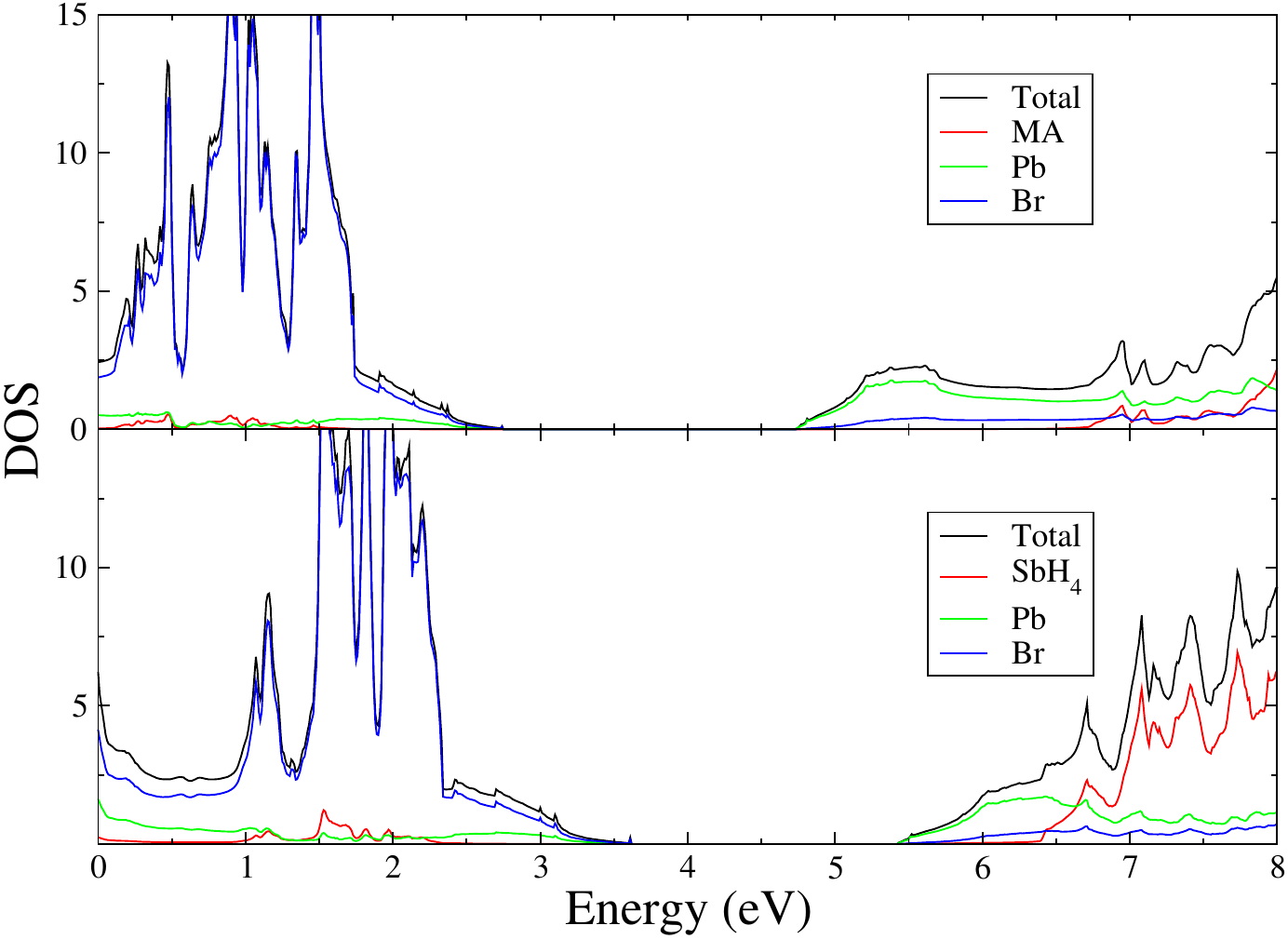}
\caption{Density of States (DOS) for MAPbBr$_3$ and SbH$_4$PbBr$_3$. The plots display the total DOS along with the individual contributions from the organic cation (MA$^+$ for MAPbBr$_3$ and SbH$_4$ for SbH$_4$PbBr$_3$), Pb, and I. Fermi level is not normalized to zero in these plots. Energy is given in electron volts (eV).}
\label{fig5}
\end{figure}

\begin{figure}[h]
\includegraphics[width=8cm]{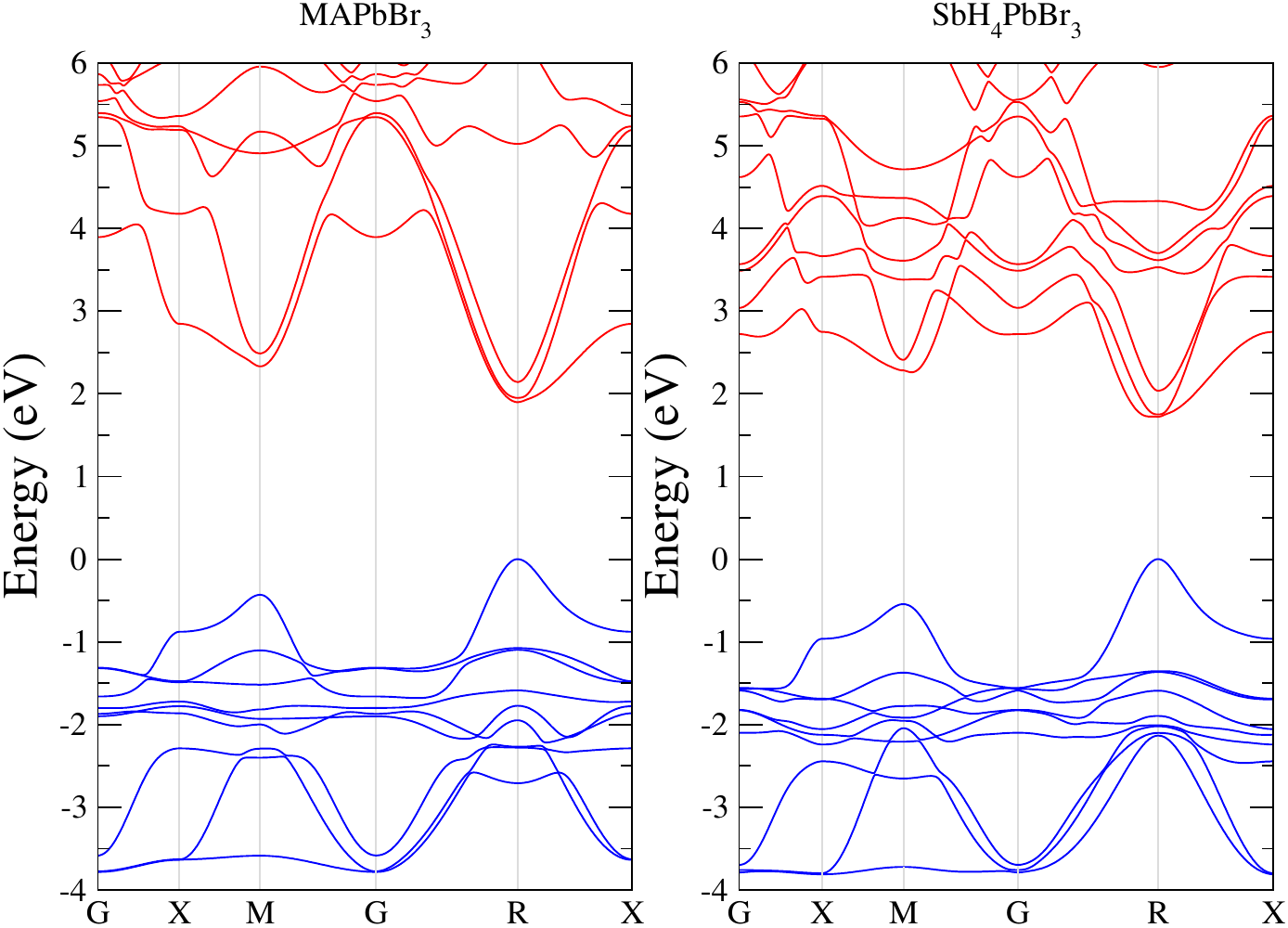}
\caption{Band structures of MAPbBr$_3$ and SbH$_4$PbBr$_3$ along high-symmetry paths in the Brillouin zone. For these band structure plots, the Fermi level is shifted to zero.}
\label{fig6}
\end{figure}

\begin{figure}[h]
\includegraphics[width=8cm]{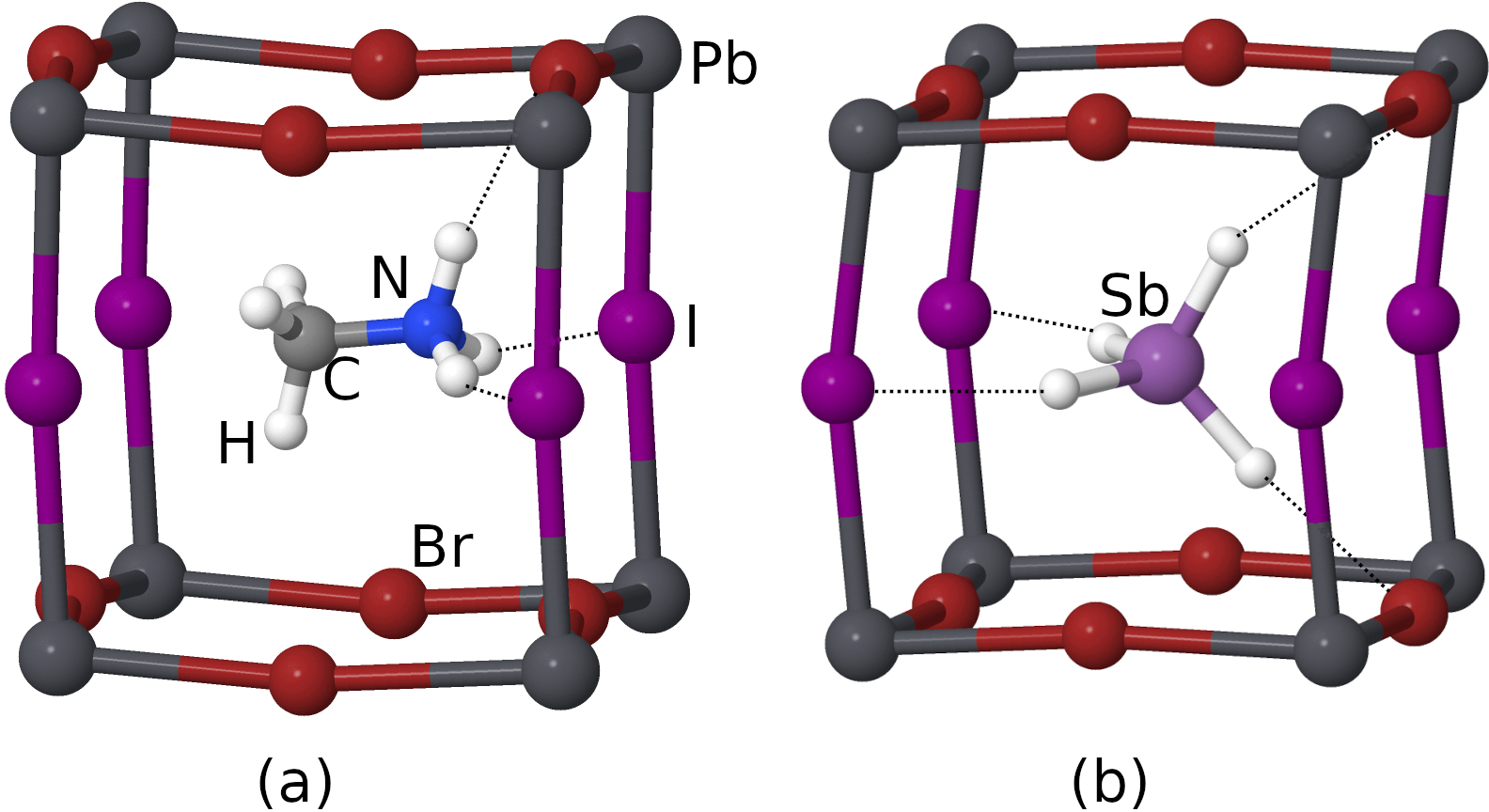}
\caption{The optimized structures of MAPbBr$_2$I and SbH$_4$PbBr$_2$I. In the inorganic lattice, gray, red and purple balls represent Pb, Br and I ions, respectively.}
\label{fig7}
\end{figure}

\begin{figure}[h]
\includegraphics[width=8cm]{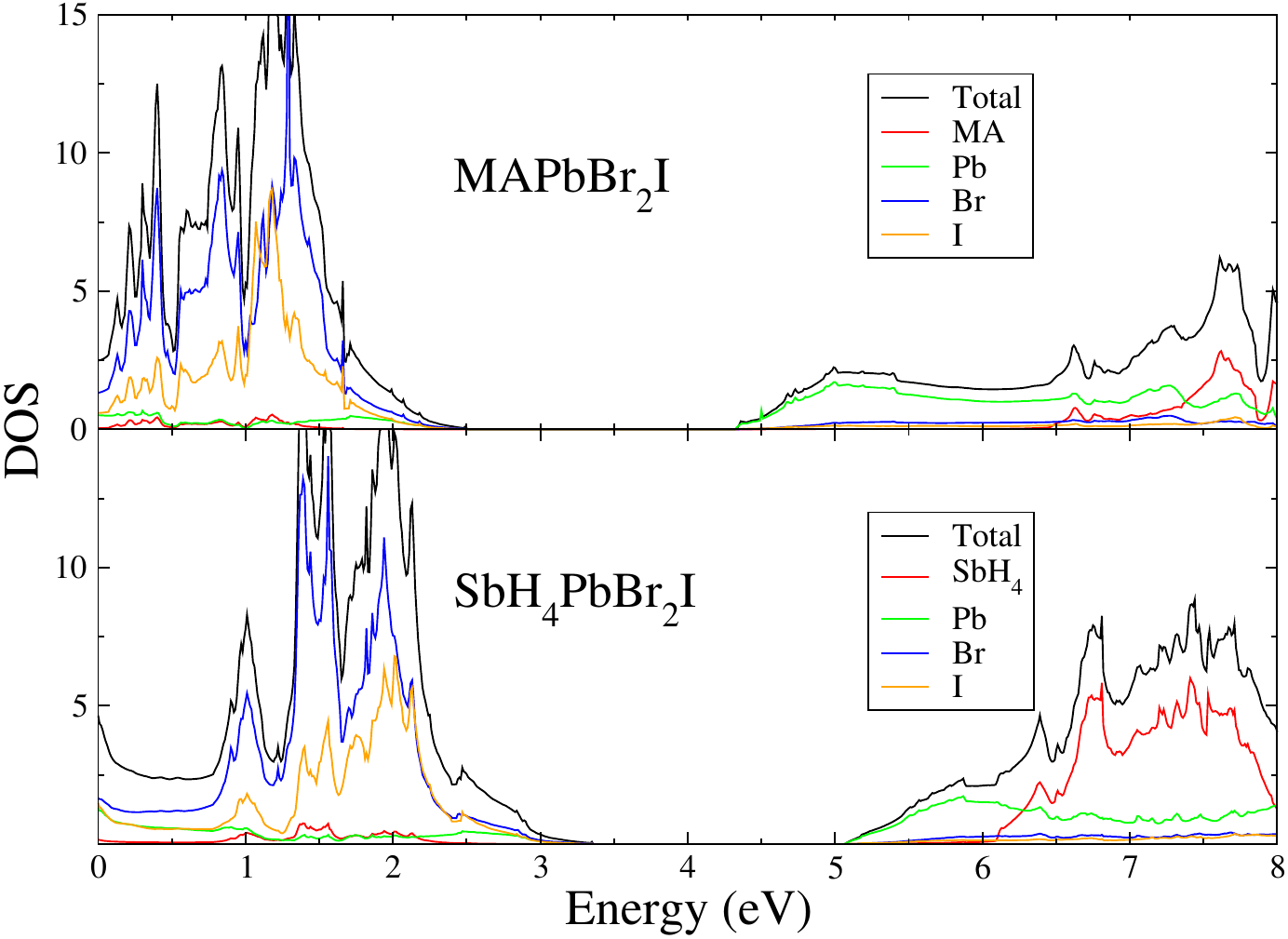}
\caption{Density of States (DOS) for MAPbBr$_2$I and SbH$_4$PbBr$_2$I. The plots display the total DOS along with the individual contributions from the organic cation (MA$^+$ for MAPbBr$_2$I and SbH$_4$ for SbH$_4$PbBr$_2$I), Pb, and I. Fermi level is not normalized to zero in these plots. Energy is given in electron volts (eV).}
\label{fig8}
\end{figure}

\begin{figure}[h]
\includegraphics[width=8cm]{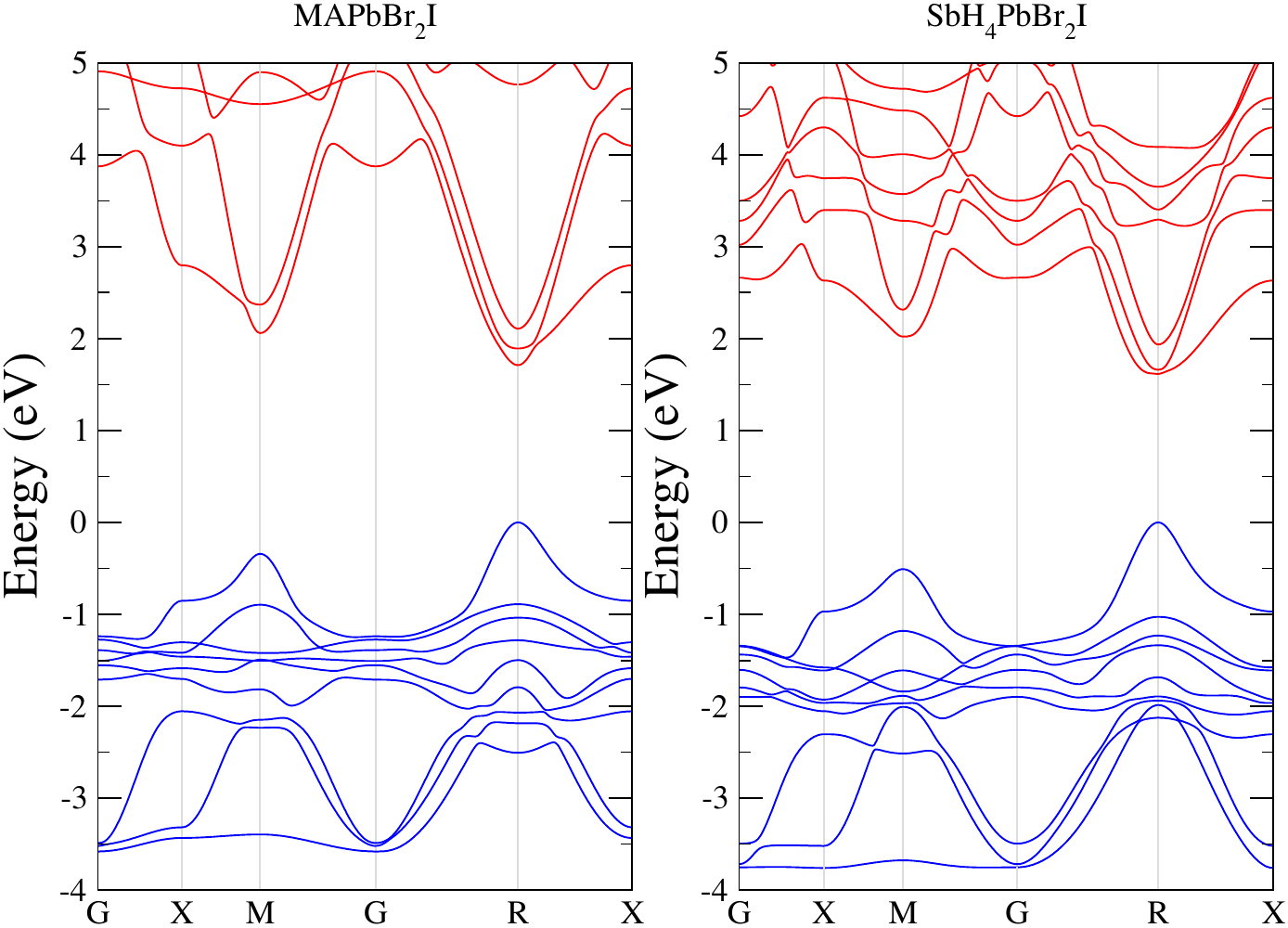}
\caption{Band structures of MAPbBr$_2$I and SbH$_4$PbBr$_2$I along high-symmetry paths in the Brillouin zone. For these band structure plots, the Fermi level is shifted to zero.}
\label{fig9}
\end{figure}

\begin{figure}[h]
\includegraphics[width=8cm]{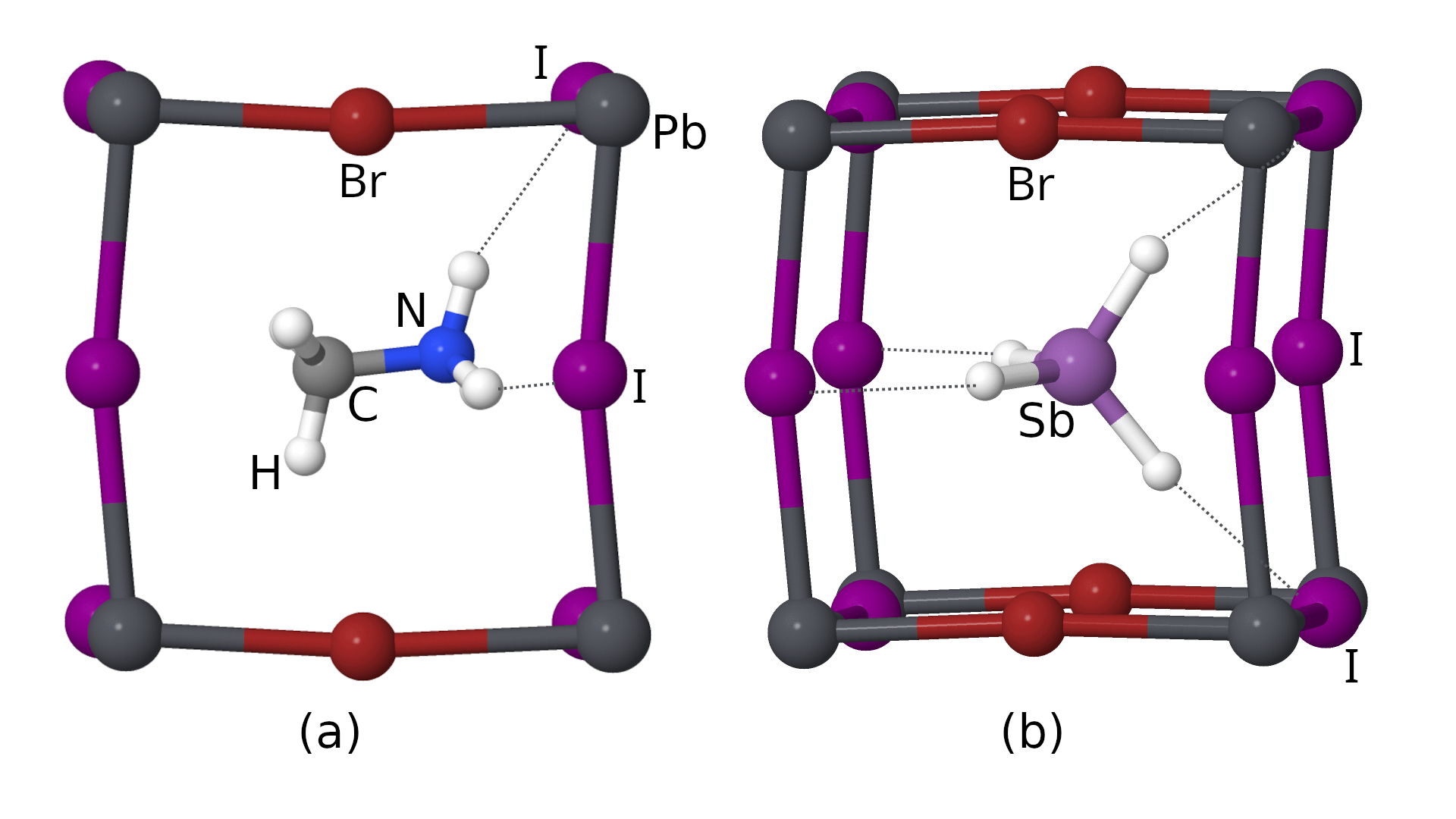}
\caption{The optimized structures of MAPbBrI$_2$ and SbH$_4$PbBrI$_2$. In the inorganic lattice, gray, red and purple balls represent Pb, Br and I ions, respectively.}
\label{fig10}
\end{figure}

\begin{figure}[h]
\includegraphics[width=8cm]{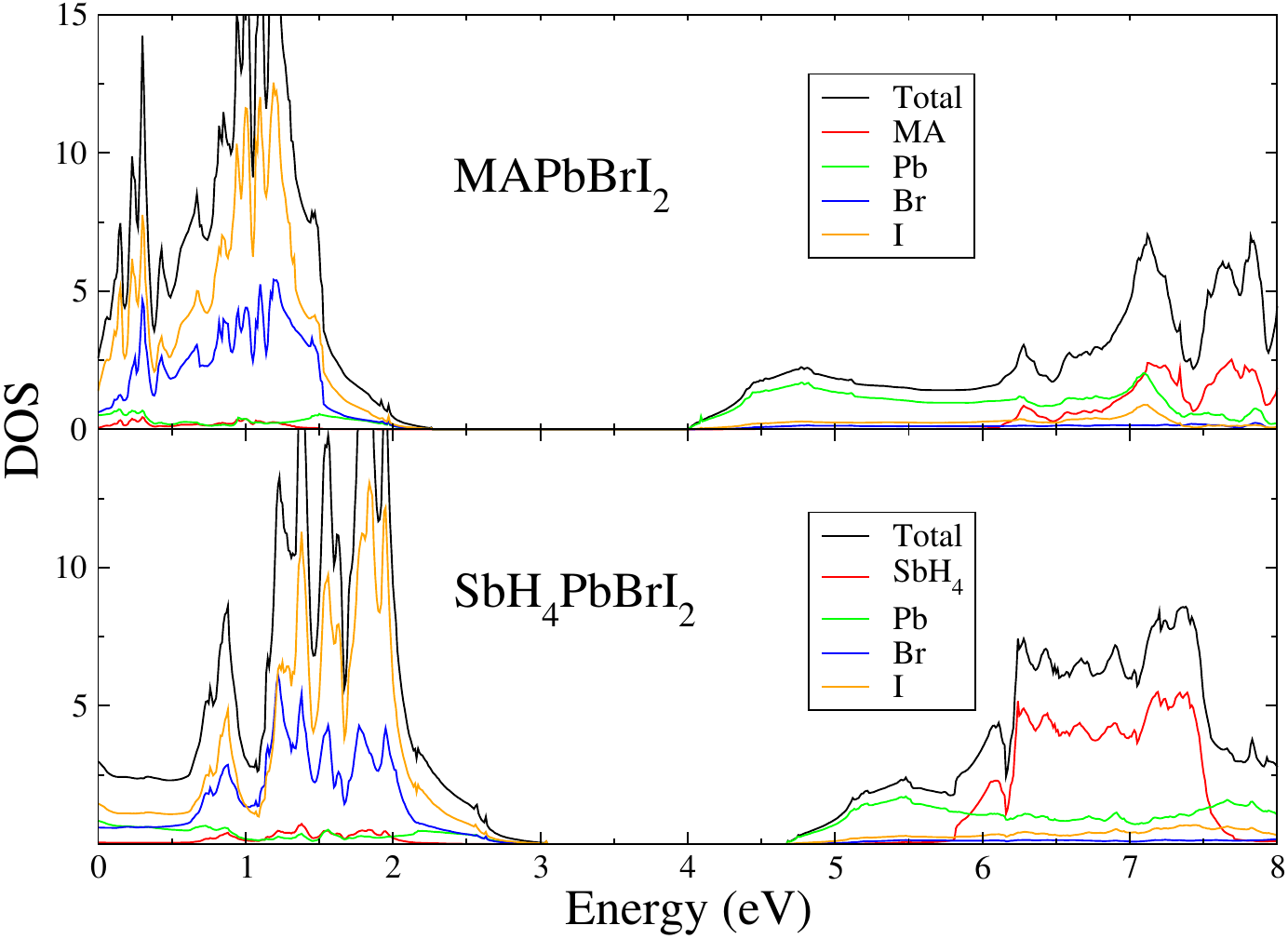}
\caption{Density of States (DOS) for MAPbBrI$_2$ and SbH$_4$PbBrI$_2$. The plots display the total DOS along with the individual contributions from the organic cation (MA$^+$ for MAPbBrI$_2$ and SbH$_4$ for SbH$_4$PbBrI$_2$), Pb, and I. Fermi level is not normalized to zero in these plots. Energy is given in electron volts (eV).}
\label{fig11}
\end{figure}

\begin{figure}[h]
\includegraphics[width=8cm]{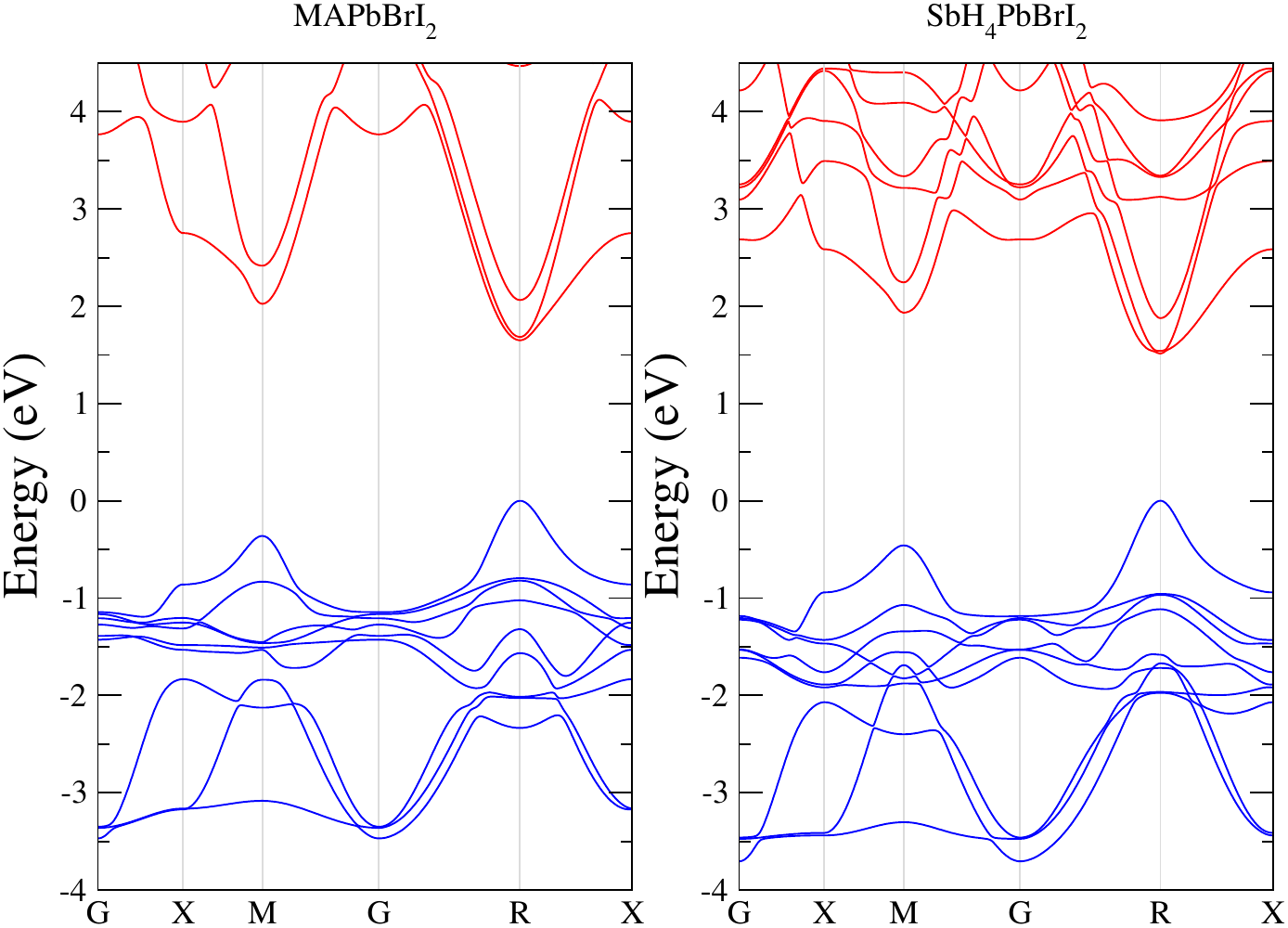}
\caption{Band structures of MAPbBrI$_2$ and MAPbBrI$_2$ along high-symmetry paths in the Brillouin zone. For these band structure plots, the Fermi level is shifted to zero.}
\label{fig12}
\end{figure}

Figure~\ref{fig2} presents the density of states (DOS) for MAPbI$_3$ and SbH$_4$PbI$_3$. In MAPbI$_3$, the valence band is predominantly composed of iodine (I) states, whereas the conduction band is mainly derived from the empty states of lead (Pb). The MA$^+$ cation contributes negligibly near the valence band maximum (VBM) and conduction band minimum (CBM). In SbH$_4$PbI$_3$, the DOS exhibits similar features; however, the SbH$_4^+$ cation appears to have a greater influence on the conduction band at higher energy levels, even though its states lie below the VBM and above the CBM. A significant difference between the two materials is observed in the relative positioning of the band edges. In SbH$_4$PbI$_3$, the VBM and CBM are shifted upward by approximately 0.75~eV and 0.56~eV, respectively, compared to MAPbI$_3$. This upward shift should also be compatible with the electron transport layers (ETLs) in perovskite solar cells. The ETL layer is important for the overall efficiency in perovskite solar cells~\cite{green2014emergence, kojima2009organometal}.

Figure~\ref{fig3} shows the calculated band structures of the two materials. Electronic transitions across the band gap are direct in both cases. The calculated band gap of MAPbI$_3$ is 1.56~eV, which agrees well with experimentally observed values between 1.55~eV and 1.60~eV~\cite{eperon2014, noh2013}. Underestimation of band gaps by standard DFT calculations is to be expected, but addition of relativistic effects due to the heavy Pb ion is likely to be the cause of this good agreement. Substitution of the MA$^+$ cation with SbH$_4^+$ causes band gap reduction. The calculated band gap of SbH$_4$PbI$_3$ is 1.37~eV, roughly 0.2~eV lower than the band gap of MAPbI$_3$. The calculated band gap is in good agreement with the 1.4~eV value obtained with GW calculations in previous theoretical works~\cite{Filip2014}. Since the absorber with perfect direct band gap near 1.4~eV in solar cells would be ideal for high-efficiency photovoltaics, such a fact means SbH$_4$PbI$_3$ is highly qualified for the photovoltaic application need of high efficiency.

The band gap energy ($E_g$) of a semiconductor is one of the critical parameters in determining the efficiency of solar cells. The Shockley-Queiser (SQ) limit prescribes that the theoretical maximum efficiency (approximately 33.7~\%) of single-junction cells is achieved at an optimum band gap of approximately 1.34~eV~\cite{shockley1961}. This is an optimum between efficient photon absorption across the solar spectrum and minimized thermalization losses. Practically, cost-efficient and stable materials such as crystalline silicon ($E_g \sim 1.1$~eV) dominate the market despite offering practical efficiencies of around 27.3\%, lower than the SQ limit~\cite{green2024}. High-efficiency III-V semiconductors such as GaAs ($E_g \sim 1.43$~eV) record efficiencies near 30\% under concentrated illumination; however, their high cost limits widespread adoption~\cite{geisz2020}. New materials, like perovskites, such as FAPbI$_3$, offer band gaps around 1.5 eV and have demonstrated high power conversion efficiencies (up to 25.8\%) with excellent stability, making them promising candidates for integration into tandem solar cell architectures that aim to surpass the single junction efficiency limit ~\cite{Min2021}. Multijunction solar cells utilizing stacks of semiconductors with progressively lower band gaps (such as InGaP/GaAs/Ge)have demonstrated record laboratory efficiencies approaching 47\% under concentrated illumination, thanks to their optimized spectral utilization across the solar spectrum~\cite{green2024, NREL2025}. However, deviations from the ideal band gap create trade-offs: smaller band gaps (less than 1.1~eV) enhance thermal losses, whereas larger band gaps (greater than 1.7~eV) decrease photocurrent generation~\cite{ruhle2017}. Thus, the ideal range of band gaps for efficient solar cells is deemed to be around 1.1--1.7~eV, successfully compromising between theoretical ideals and practical limitations.

The calculated band gap of 1.37~eV in this work for SbH$_4$PbI$_3$ signifies that SbH$_4$PbI$_3$ could be a more effective perovskite material than MAPbI$_3$ for band gap utility in solar cells. However, it is acknowledged that perovskites composed of organic cations and PbI$_3^-$ structures are generally very unstable. Therefore, a strategic solution is required to achieve a balance between keeping the band gap at an optimal value and enhancing the stability of the material simultaneously. In addition to iodide perovskites, bromide-based and mixed-halide architectures have also been extensively researched as outstanding alternatives to satisfy the conditions of achieving photovoltaic efficiency with long-term stability. As a demonstration, PbBr$_3^-$ is generally more thermally and humid-stable than PbI$_3^-$~\cite{noh2013}. Substitution of I$^-$ ions with Br$^-$ ions results in the decrease of lattice parameters due to the smaller radius of Br$^-$ ion, which has the potential to increase stability. The calculated structural data are listed in Table~\ref{Table1}.

The contrast between MAPbI$_3$ and MAPbBr$_3$ shows dramatic structural and electronic modifications due to the substitution of iodine with bromine. For MAPbBr$_3$, the unit cell parameters ($a = 5.96$~\AA, $b = 5.96$~\AA, $c = 5.99$~\AA) are considerably smaller than those for MAPbI$_3$ ($a = 6.36$~\AA, $b = 6.33$~\AA, $c = 6.37$~\AA). This contraction of the perovskite lattice is consistent with the smaller ionic radius of Br$^-$ compared to I$^-$. Additionally, the 3.03~\AA\ Pb--Br bond length is shorter than that of the 3.22~\AA\ Pb--I bond, and the 2.47~\AA\ H--Br bond length is shorter than the 2.71~\AA\ H--I bond length. These differences indicate a tighter hydrogen-halide interaction and a tighter crystal structure for the bromide perovskite.

Electronically, MAPbBr$_3$ possesses a larger valence band maximum (VBM) of 2.8141~eV over 2.1799~eV for MAPbI$_3$, and the conduction band minimum (CBM) is larger as well (4.7151~eV for MAPbBr$_3$ vs 3.7387~eV for MAPbI$_3$). Consequently, the band gap is enlarged from 1.5588~eV in MAPbI$_3$ to 1.9010~eV in MAPbBr$_3$. Moreover, the electron effective mass is increased from 0.254 to 0.268 and the hole effective mass is altered from 0.232 in MAPbI$_3$ to 0.222 in MAPbBr$_3$, which means bromine substitution influences both the band alignment and charge carrier dynamics, and hence the material's optoelectronic properties.

A similar trend is observed in the comparison of SbH$_4$PbI$_3$ and SbH$_4$PbBr$_3$. The lattice parameters for SbH$_4$PbBr$_3$ ($a = 5.84$~\AA, $b = 5.91$~\AA, $c = 5.96$~\AA) are distinctly smaller than those of SbH$_4$PbI$_3$ ($a = 6.23$~\AA, $b = 6.30$~\AA, $c = 6.36$~\AA), further confirming the contraction due to the smaller size of Br$^-$. The Pb--Br bond length in SbH$_4$PbBr$_3$ is 2.96~\AA, which is shorter than the Pb--I bond length of 3.14~\AA\ in SbH$_4$PbI$_3$, and the H--Br bond length (2.75~\AA) is also less than the H--I bond length (2.94~\AA). Electronics of SbH$_4$PbBr$_3$ are similarly affected: its VBM is 3.6940~eV compared to 2.9328~eV in SbH$_4$PbI$_3$, and the CBM rises from 4.3003~eV in SbH$_4$PbI$_3$ to 5.4169~eV in SbH$_4$PbBr$_3$. This results in an increased band gap from 1.3675~eV in SbH$_4$PbI$_3$ to 1.7229~eV in SbH$_4$PbBr$_3$, indicating a stronger quantum confinement effect with bromine incorporation. Furthermore, the effective masses of both electrons and holes increase, with the electron effective mass rising from 0.215 to 0.289 and the hole effective mass increasing from 0.164 to 0.182. These variations emphasize that bromine substitution not only alters the band alignment but also modulates the charge carrier transport properties, which can significantly affect the material’s optoelectronic performance.

Figure~\ref{fig4} presents the optimized structures of MAPbBr$_3$ and SbH$_4$PbBr$_3$. When comparing the MAPbBr$_3$ and SbH$_4$PbBr$_3$ structures, the lattice parameters are very similar, with only a slight contraction in the $a$ parameter observed upon substituting MA$^+$ with SbH$_4^+$ (Table~\ref{Table1}). Such subtle differences in hydrogen-bonding interactions, where MA$^+$ forms N--H$\cdots$Br bonds and SbH$_4^+$ establishes additional H$\cdots$Br contacts, are also reflected in the electronic structures. The DOS of both materials shows a valence band predominantly composed of Br$^-$ states and a conduction band dominated by Pb states (Figure~\ref{fig5}). As can be seen in Figure~\ref{fig6}, the minimum band gap is located at the R symmetry point, and the nature of the transition is direct in both structures. Replacing the MA$^+$ cation with the SbH$_4^+$ cation in the PbI$_3^-$ inorganic framework narrows the band gap, resulting in a more suitable value of 1.36~eV. However, when the PbI$_3^-$ anion is replaced with the PbBr$_3^-$ anion to enhance stability, the band gap increases from 1.36~eV to 1.72~eV, as shown in Table~\ref{table2}. This indicates that a fine-tuning process is necessary to optimize both electronic properties and stability. A suitable strategy would be the gradual substitution of Br$^-$ ions with I$^-$ ions.

The optimized structure of MAPbBr$_2$I and SbH$_4$PbBr$_2$I are presented in Figure~\ref{fig7} and relative data for such structure are in Table~\ref{Table1}. Lattice parameters $a = 5.96$, $b = 5.96$, and $c = 5.99$~\AA, Pb--Br and H--Br bond lengths 3.03~\AA\ and 2.47~\AA, respectively. Substitution of a Br$^-$ by an I$^-$ to form MAPbBr$_2$I increases the $c$-axis to 6.362~\AA\ and causes $a$ and $b$ to be little changed. The substitution adds a Pb--I bond of 3.20~\AA\ and an H--I bond of 2.60~\AA, as might be expected of I$^-$'s increased ionic radius and localized structural impact. The same holds true for SbH$_4$PbBr$_3$, whose lattice parameters are $a = 5.84$, $b = 5.91$, and $c = 5.96$~\AA, and whose Pb--Br and H--Br bond distances are 2.96~\AA\ and 2.75~\AA, respectively. If one Br$^-$ is substituted by an I$^-$ to form SbH$_4$PbBr$_2$I, the $c$ parameter increases to 6.287~\AA, while $a$ and $b$ increase very slightly ($a = 5.883$~\AA, $b = 5.936$~\AA). The Pb--I bond length in SbH$_4$PbBr$_2$I is also slightly shorter (3.15~\AA) than in MAPbBr$_2$I (3.20~\AA), while the Pb--Br bond is close to its original value (2.97~\AA\ versus 3.03~\AA\ in MAPbBr$_2$I). The hydrogen-bonding arrangement is also severely altered, with H--I and H--Br bond lengths increasing to 2.86~\AA\ and 2.82~\AA, respectively, in SbH$_4$PbBr$_2$I, from the respective 2.60~\AA\ and 2.41~\AA\ in MAPbBr$_2$I. Both complexes preserve the perovskite structure, with Pb coordinated to the halide anions.

The electronic properties of the perovskite materials MAPbBr$_3$, MAPbBr$_2$I, SbH$_4$PbBr$_3$, and SbH$_4$PbBr$_2$I have been examined with particular interest in the effects of halide substitution and A-site cation change. Within the MAPbBr$_3$ framework, iodide substitution makes both the valence band maximum and the conduction band minimum shift downwards, reducing the band gap from $1.9010\ \text{eV}$ to $1.7126\ \text{eV}$ (Table~\ref{table2}). This is because of the lower electronegativity and bigger ionic radius of the iodide ion in comparison to bromide, which alters orbital interaction and energy level alignment. Furthermore, the effective masses of charge carriers are reduced in MAPbBr$_2$I, with the electron effective mass decreasing from $0.268$ to $0.170$ and that of the hole from $0.222$ to $0.194$, thereby suggesting higher carrier mobility beneficial for optoelectronic applications.

A comparable trend is observed in the SbH$_4$PbBr$_3$ and SbH$_4$PbBr$_2$I structures. For SbH$_4$PbBr$_3$, the valence band and conduction band edges are located at $3.6940\ \text{eV}$ and $5.4169\ \text{eV}$, respectively, resulting in a band gap of $1.7229\ \text{eV}$ (Table~\ref{table2}). In SbH$_4$PbBr$_2$I, the corresponding band edges shift to $3.4400\ \text{eV}$ and $5.0579\ \text{eV}$, yielding a reduced band gap of $1.6179\ \text{eV}$. However, in this system the electron effective mass increases from $0.289$ to $0.360$, while the hole effective mass decreases slightly from $0.182$ to $0.175$. These changes indicate that the iodide substitution produces a more complex impact on the charge carrier dynamics by adversely affecting electron mobility while modestly enhancing hole transport.

The comparative analysis of MAPbBr$_2$I and SbH$_4$PbBr$_2$I perovskites clearly illustrates the crucial role of the A‐site cation in the electronic structure. The DOS plot for the MAPbBr$_2$I and SbH$_4$PbBr$_2$I structures is presented in Figure~\ref{fig8}, while the corresponding electronic band structure is illustrated in Figure~\ref{fig9}. The valence band is primarily composed of halogen (Br$^-$, I$^-$) orbitals, while the conduction band edge is composed of empty Pb orbitals. However, when looking at the degree to which the contribution from MA$^+$ and SbH$_4^+$ cations is made, SbH$_4^+$ contributes proportionately more, particularly in determining the minimum conduction band (CBM). In MAPbBr$_2$I, organic MA$^+$ cation creates lower energy valence and conduction bands, whereas the inorganic SbH$_4^+$ cation in SbH$_4$PbBr$_2$I engages more with the nearby halide ions and shifts these bands to higher energy. While SbH$_4$PbBr$_2$I possesses a slightly narrower band gap than MAPbBr$_2$I, as can be seen from Table~\ref{table2}, band edge position changes and the associated effective mass changes result in extreme changes in both optical absorption and charge transport properties.

In the case of the MA-based compounds, MAPbBr$_2$I exhibits lattice parameters of $a = 5.952$~\AA, $b = 5.989$~\AA, and $c = 6.362$~\AA, whereas MAPbBrI$_2$ shows slightly increased values of $a = 5.947$~\AA, $b = 6.359$~\AA, and $c = 6.407$~\AA. This moderate expansion, particularly along the $c$-axis, is attributed to the larger ionic radius of I$^-$ relative to Br$^-$. The optimized structures of MAPbBrI$_2$ and SbH$_4$PbBrI$_2$ are presented in Figure~\ref{fig10}, and the corresponding data for these structures are listed in Table~\ref{Table1}. The bond lengths further corroborate these findings; MAPbBr$_2$I has Pb--I and Pb--Br bond lengths of 3.20~\AA\ and 3.03~\AA, respectively, while in MAPbBrI$_2$ these bonds are slightly elongated to 3.22~\AA\ and 3.06~\AA. The increase in the number of I$^-$ ions in MAPbBrI$_2$ also affects the hydrogen bond network. The H--I bond length, which is 2.60 ~\AA\ in MAPbBr$_2$I, increases to 2.64~\AA\ with the increase in the number of I$^-$ ions in the lattice. The optimized structures shown in Figure~\ref{fig10} are the structures with the lowest energy among the tested configurations. In these structures, hydrogens prefer to bond with the I ion when compared to the Br$^-$ ion. As can be seen in the structures obtained as a result of the calculations in Figure~\ref{fig10}, no H--Br bond is formed. However, there is not much energy difference between the compared alternative configurations. Accordingly, the orientations of the cation molecules may be sensitive to heat. 

In the SbH$_4$-based perovskites, SbH$_4$PbBr$_2$I displays lattice constants of $a = 5.883$~\AA, $b = 5.936$~\AA, and $c = 6.287$~\AA, while SbH$_4$PbBrI$_2$ shows a modest expansion to $a = 5.879$~\AA, $b = 6.310$~\AA, and $c = 6.311$~\AA, particularly along the $b$- and $c$-axes. Correspondingly, the Pb--I bond length in SbH$_4$PbBr$_2$I (3.15~\AA) increases slightly to 3.17~\AA\ in SbH$_4$PbBrI$_2$, whereas the Pb--Br bond length decreases from 2.97~\AA\ to 2.93~\AA. The hydrogen bonding environment is altered as well, with SbH$_4$PbBr$_2$I exhibiting H--I and H--Br bond lengths of 2.86~\AA\ and 2.82~\AA, respectively, and SbH$_4$PbBrI$_2$ showing an H--I bond length of 2.88~\AA. Moreover, a direct comparison between MAPbBrI$_2$ and SbH$_4$PbBrI$_2$ highlights the impact of A-site cation substitution: the replacement of the bulky organic MA$^+$ cation with the more compact SbH$_4^+$ cation results in a slight contraction of the lattice and in shorter Pb--halide bonds. The hydrogen bonding is also distinctly modified, with MAPbBrI$_2$ showing an H--I bond of 2.64~\AA\ and lacking an H--Br bond, whereas in SbH$_4$PbBrI$_2$ the H--I bond length is increased to 2.88~\AA. These observations suggest that the introduction of the inorganic SbH$_4^+$ cation yields a denser and more rigid perovskite framework, which may enhance the material’s thermal and chemical stability.

Crossing over to the electronic characteristics, Br$^-$ substitution by I$^-$ fundamentally affects the band structure. The comparative DOS pattern of the MAPbBrI$_2$ and SbH$_4$PbBrI$_2$ lattices is provided in Figure~\ref{fig11}, and their corresponding electronic band structure is presented in Figure~\ref{fig12}. For the MA-based perovskites, the VBM of MAPbBr$_2$I is 2.5681~eV and that of the CBM is 4.2807~eV, resulting in a band gap of 1.7126~eV. For MAPbBrI$_2$, the VBM turns out to be 2.3287~eV and the CBM turns out to be 3.9798~eV, resulting in the band gap further reducing to 1.6511~eV (Table~\ref{table2}). This narrowing of band gap is primarily caused by lower electronegativity and higher ionic radius of iodine, which result in shifting of the energy levels and allow greater absorption of light in longer wavelength region. In addition, carrier dynamics of charges are modified; MAPbBr$_2$I possesses an effective mass of a hole and electron of 0.170 and 0.194, whereas those of MAPbBrI$_2$ are up to 0.239 and 0.198, meaning electron mobility can have a decrease.

In SbH$_4$-based perovskites, the trend is the same. SbH$_4$PbBr$_2$I has a VBM of 3.4400~eV and a CBM of 5.0579~eV, which translates into a band gap of 1.6179~eV. On substituting iodide in place of addition in SbH$_4$PbBrI$_2$, the VBM decreases to 3.1270~eV and the CBM to 4.6416~eV, reducing the band gap to 1.5146~eV. Remarkably, the effective electron mass of 0.360 in SbH$_4$PbBr$_2$I decreases markedly to 0.147 in SbH$_4$PbBrI$_2$, although the hole effective mass neither practically varies (0.175 vs. 0.158). All these suggest SbH$_4$PbBrI$_2$ to exhibit improved mobility and charge transport with improved qualities over others for optoelectronic application.

The combined structural and electronic studies demonstrate that substitution of Br$^-$ with I$^-$ induces a moderate lattice expansion, Pb--I and H--I bond lengthening, and alteration of the network of hydrogen bonding. Concurrently, substitution of the organic MA$^+$ cation with the inorganic SbH$_4^+$ cation results in a higher density of perovskite framework and shorter Pb--halide bonds along with significant effects on the electronic density of states. These modifications result in smaller band gaps and modified effective masses of charge carriers and hence impact the optical absorption and charge transport properties of the materials. This tunability of both structure and electronic properties is suggestive of being able to tailor these perovskites for improved performance in photovoltaic and other optoelectronic applications.

\section{CONCLUSION}

In this study, the structural and electronic properties of hybrid perovskite materials, namely the comparisons between MAPbBr$_x$I$_{1-x}$ and SbH$_4$PbBr$_x$I$_{1-x}$ systems, are explicitly investigated by first-principles DFT calculations with van der Waals corrections. The results show that substituting the conventional MA$^+$ cation with SbH$_4^+$ can improve the electronic properties and tighten the structure in a way that increases the structural stability.

The addition of the SbH$_4^+$ cation can provide lower lattice parameters and shorter Pb--halogen bond lengths, leading to the production of more stable and denser crystals. Substitution with SbH$_4^+$ also results in an increase in the valence band maximum and a reduced band gap. Of the investigated compounds, SbH$_4$PbI$_3$ is particularly promising with an optimum band gap of approximately 1.37~eV, which is quite close to the ideal band gap for single-junction solar cells. The optimum band gap, increased charge carrier mobilities, and reduced recombination losses implied by effective mass calculations suggest that SbH$_4$PbI$_3$ could be a good alternative material candidate to conventional MAPbI$_3$ for high-efficiency photovoltaic applications.

It is known that the incorporation of Br$^-$ ions into the inorganic cage of PbI$_3^-$ increases structural stability. The calculations performed in this study indicate that SbH$_4$PbBrI$_2$ is one of the important candidates, as it can provide additional structural stability to the electronic values of iodide by incorporating bromide. Electronically, SbH$_4$PbBrI$_2$ has a suitable band gap of approximately 1.51~eV, balanced between MAPbI$_3$ and SbH$_4$PbI$_3$, and has suitable charge carrier dynamics. Therefore, SbH$_4$PbBrI$_2$ is a strong candidate for photovoltaic devices with a compromise between structural stability and optimal electronic properties.

Moreover, controlled halide substitution, i.e., replacing iodide ions with bromide ions, is a valuable method to achieve maximum material stability at the expense of electronic performance. However, adding Br$^-$ to the MAPbI$_3$ structure increases the band gap from about 1.55~eV to 1.90~eV. In this case, the stability increases but the efficiency decreases. In SbH$_4$PbBrI$_2$, the SbH$_4^+$ cation can tolerate the band gap to lower levels. For example, the band gap of SbH$_4$PbBrI$_2$ is about 1.51~eV.

In conclusion, the extensive theoretical investigation presented here strongly supports the use of SbH$_4^+$ substitution as a practical method to develop perovskite materials with improved stability and optimal electronic properties. In particular, SbH$_4$PbI$_3$ and SbH$_4$PbBrI$_2$ are found to be leading candidates for real solar energy applications, worthy of future experimental validation to fully realize their potential and make them commercially acceptable.

\begin{acknowledgments}
The numerical calculations reported in this paper were fully performed at TUBITAK ULAKBIM, High Performance and Grid Computing Center (TRUBA resources).
\end{acknowledgments}

\bibliographystyle{unsrt}
\bibliography{sbh4pbbri}

\end{document}